\documentclass[aps,prd,preprint,
superscriptaddress,tightenlines,nofootinbib,showpacs]{revtex4}
\usepackage{graphicx}
\usepackage{dcolumn}

\newcommand{\BBbar}{B\overline{B}}

\newcommand{\BBM}{B^0 \overline{B}^0}

\newcommand{\ep}{e/\overline p}
\newcommand{\ctt}{cos(\theta)}

\newcommand{\pepp}{\vert \vec P_{\overline p} \vert + \vert \vec P_{e} \vert}

\newcommand{\btoppen} {B^-\rightarrow p\overline p e^-\overline \nu_{e}}

\newcommand{\btoxpen} {B \rightarrow \overline p e^-\overline \nu_{e} X}

\newcommand{\btolcpen} {B^-\rightarrow\Lambda_{c}^+\overline p e^-\overline \nu_{e}}
\newcommand{\btoalcpen} {B^-\rightarrow\Lambda_{c}^+ \overline p e^- \overline \nu_{e}}

\newcommand{\btou} {b \rightarrow u}
\newcommand{\btoc} {b \rightarrow c}
\newcommand{\GeVc} {\mbox{\rm GeV/$c$}}

\begin{document}


\preprint{CLNS 03-1820}
\preprint{CLEO 03-06}

\title{Search for $\btoxpen$ decay using a 
partial reconstruction method}

\author{N.~E.~Adam}
\author{J.~P.~Alexander}
\author{K.~Berkelman}
\author{V.~Boisvert}
\author{D.~G.~Cassel}
\author{P.~S.~Drell}
\author{J.~E.~Duboscq}
\author{K.~M.~Ecklund}
\author{R.~Ehrlich}
\author{R.~S.~Galik}
\author{L.~Gibbons}
\author{B.~Gittelman}
\author{S.~W.~Gray}
\author{D.~L.~Hartill}
\author{B.~K.~Heltsley}
\author{L.~Hsu}
\author{C.~D.~Jones}
\author{J.~Kandaswamy}
\author{D.~L.~Kreinick}
\author{A.~Magerkurth}
\author{H.~Mahlke-Kr\"uger}
\author{T.~O.~Meyer}
\author{N.~B.~Mistry}
\author{J.~R.~Patterson}
\author{D.~Peterson}
\author{J.~Pivarski}
\author{S.~J.~Richichi}
\author{D.~Riley}
\author{A.~J.~Sadoff}
\author{H.~Schwarthoff}
\author{M.~R.~Shepherd}
\author{J.~G.~Thayer}
\author{D.~Urner}
\author{T.~Wilksen}
\author{A.~Warburton}
\author{M.~Weinberger}
\affiliation{Cornell University, Ithaca, New York 14853}
% ...

\date{\today}

\begin{abstract}
Using data collected on the $\Upsilon(4S)$ resonance and the nearby continuum  
by the CLEO detector at the Cornell Electron Storage Ring, 
we have searched for the semileptonic decay of $B$ mesons to 
$e \overline p$ inclusive 
final states. We obtain 
an upper limit for $b\rightarrow c$ decays of 
${\cal B} (B\rightarrow\overline{p}e^-\overline \nu_{e}X)<5.9\times 10^{-4}$.
For the $b \rightarrow u$ decay, we find an upper limit of 
${\cal B} (\btoppen)<1.2\times 10^{-3}$ based on a V-A model, 
while a phase space model gives an upper limit of 
${\cal B} (\btoppen)<5.2\times 10^{-3}$.
All upper limits are measured at the $90\%$ confidence level. 
 
\end{abstract}
\maketitle



\section{Introduction}

Semileptonic decays play a prominent role in $B$ physics, because they are 
simple to understand theoretically and have been used to find 
$\BBM$ mixing~\cite{mixing} 
and the values of the CKM matrix elements: $V_{cb}$~\cite{vcb} and 
$V_{ub}$~\cite{b2u}. 

For many years there have been some mysteries in the 
$B$ meson semileptonic decays. 
For example,
the measured semileptonic branching fraction of 
$B$ mesons~\cite{argus,oldcleo} is about 2\% lower absolute (20\% relative) 
than theoretical predictions~\cite{theory1}.   
Recently, there has been some  
progress made on both the experimental and theoretical 
fronts~\cite{PDG,thorn-cleo,pro-ex,pro-th}, 
which gives values in better agreement with each other.  
More measurements are needed to 
improve the existing results as well as to precisely test the new theoretical 
calculations.

The majority of semileptonic $B$ decays appear to 
proceed with single mesons accompanying the lepton-antineutrino pair.  
There is no experimental 
evidence for baryons in semileptonic $B$ decay.
Therefore, in this paper, we will focus on 
the search for these decay modes.  
Baryon production in $B$ meson semileptonic decays requires the ``popping'' 
of two quark-antiquark pairs from the vacuum. For instance, 
in a $B^-$ decay, the quark content of the
baryons will be 
$(cud)(\overline{uud})$ when $b\rightarrow c$,
or $(uud) (\overline {uud})$ when $b\rightarrow u$. 
The decay mode with the lightest mass 
$b\rightarrow c$ final state including a proton 
would be $\btolcpen$. Other higher mass hadronic resonances could also 
contribute to semileptonic baryon decays with a final state having an 
electron and an antiproton.  There is little guidance for 
the probable mix of states that might be 
available so we choose a model with a mixture 
of modes to study $b\rightarrow c$ decays.  For $b\rightarrow u$ decays, 
the lightest mass final state would be either 
$\btoppen$ or $\overline{B}^0\rightarrow p\overline{n}e^-\overline{\nu}_e$.  
There is a large group of higher resonances possible.  We choose to study only 
the $\btoppen$ state in our $b\rightarrow u$ studies.

A previous CLEO~II measurement of the decay $\btoalcpen$ 
employed full reconstruction for
$\Lambda_{c}^+ \rightarrow p K^- \pi^+$~\cite{lin}. 
That analysis yielded an upper limit of
\begin{displaymath}
	\frac{{\cal B} (\btoalcpen\ )} {{\cal B} (\overline{B}\rightarrow 
	\Lambda_{c}^+\overline p\ X)}<0.04 \quad (\textrm{$C.L.=90\%$}).  
\end{displaymath}
This implies ${\cal B} (\btoalcpen)<1.7\times 10^{-3} \quad (C.L.=90\%)$ using 
the PDG value for $\overline{B}\rightarrow 
	\Lambda_{c}^+\overline p\ X$~\cite{PDG}.  
There is also an upper limit on the 
inclusive rate of 
${\cal B} (B\rightarrow \overline p e^- \nu_{e} X)<1.6\times10^{-3}~
(C.L.=90\%)~$\cite{arg2} from ARGUS.  There are no measurements of the 
$\btoppen$ decay.  

We perform partial reconstruction of the decay 
$B\rightarrow\overline{p}e^-\overline{\nu_e}X$, by identifying events 
with an $e^-$ and $\overline{p}$ emerging promptly from the $B$ and examining 
the angular distributions between them.\footnote{Throughout this paper, 
charge conjugate states are implied.}    
Muons are not used in this analysis because 
they are only well-identified above 1.4~$\GeVc$ momentum.  
Few signal leptons are expected at such momenta. 

In Section II, we describe the data sample and event selection.
The event selection criteria are tailored to search for the decay $\btoalcpen$.
We discuss the angular 
distribution of the signal and main sources of 
backgrounds in Section III. 
Section IV describes how we fit the data distribution for the 
$b\rightarrow c$ modes. 
In Section V, we discuss 
the analysis for $\btoppen$. 
The last section summarizes our results.
 

\section{Data Sample and Event Selection}

The analysis described here is based on the data 
recorded with the CLEO detector at the Cornell Electron
Storage Ring (CESR).
The CLEO detector~\cite{CLEODetector} is a general 
purpose detector that provides charged 
particle tracking, precision electromagnetic calorimetry, 
charged particle identification and
muon detection. Charged particle detection over 
95\% of the solid angle is achieved by tracking
devices in two different configurations. In the first 
configuration (CLEO~II), tracking is provided by three 
concentric wire chambers while in the 
second configuration (CLEO~II.V), the innermost wire 
chamber is replaced by a precision 
three-layer silicon vertex detector~\cite{SVXDetector} and the 
drift chamber gas was changed from 50-50\% $Ar-C_2H_6$ to 
60-40\% $He-C_3H_8$.    
Energy loss ($dE/dx$) in the outer drift chamber and hits in the time 
of flight system just beyond it provide information 
on particle identification. 
Photon and electron showers are detected over $98\%$ of $4\pi$
steradians in an array of 7800 CsI scintillation counters.  The 
electromagnetic energy resolution is found to be 
$\delta E/E = 0.0035/E^{0.75} +0.019-0.001E$ ($E$ 
in GeV) in the central region, corresponding to the polar angle of a
track's momentum vector with respect to the \textit{z} axis (beam line), 
$45^{0}<\theta_{dip}<135^0$. A magnetic field of 1.5 T is provided by a 
superconducting coil which surrounds the calorimeter and tracking 
chambers.  

A total integrated luminosity of 9.1 {\rm fb$^{-1}$} was collected by the 
CLEO~II and CLEO~II.V configurations  
at the center-of-mass energy corresponding to the 
$\Upsilon(4S)$, corresponding to $(9.7 \pm 0.2)\times 10^6$
$\BBbar$ pairs.  An additional integrated luminosity of 4.6 {\rm fb$^{-1}$} 
taken at energies 60 MeV below the $\BBbar$ threshold provides an 
estimate of the continuum background events due to $e^+e^- \rightarrow 
q \overline q$, where $q = u,d,s,c$. 

All events considered pass the standard CLEO 
hadronic event criteria, which require at least 3 well-reconstructed 
charged tracks, a total visible energy of at least 15\% of the 
center of mass energy and 
an event vertex consistent with the known $e^+e^-$ interaction point.
In order to remove $e^+e^-\rightarrow q\overline{q}$ continuum contributions, 
the ratio of the second to zeroth Fox-Wolfram moments~\cite{r2gl} 
is required to be less than 0.35.  

Charged electron and antiproton candidates are 
selected from tracks that are well-reconstructed,
and not identified as a muon.  We accept only those 
charged tracks that are observed in the barrel region of the 
detector, which corresponds to 
$\vert$ cos($\theta_{dip}$) $\vert < 0.7071$.  
Electrons with momenta 
between 0.6~$\GeVc$ and 1.5~$\GeVc$ are 
identified by requiring that 
the ratio of their energy deposited in the CsI calorimeter and their 
momentum measured in the tracking system be close to unity and that 
the ionization 
energy loss measured by the tracking system be consistent with the 
electron hypothesis.  The ratio of the log of the likelihood for the 
electron hypothesis to that 
for a hadron is required to be greater than 3.  
Electrons within the fiducial volume in this momentum range are 
identified with an efficiency of $\sim $94\%.  
Where possible, electrons from $\gamma$ conversion, $\pi^0$ Dalitz decays, 
and $J/\psi$ decays are explicitly vetoed by cuts on the 
appropriate invariant mass 
distribution.  Antiprotons with momentum 
between 0.2~$\GeVc$ and 1.5~$\GeVc$ are  
identified using the combined information from $dE/dx$ and TOF measurements. 
Antiproton candidates must lie within 3.0 standard 
deviations ($\sigma$) of the antiproton hypothesis and outside of 
2.0 $\sigma$ for each of the kaon and pion hypotheses.

To suppress correlated background (see below) 
of the $B$, we perform a primary vertex ($e^+e^-$
interaction point) constrained fit to the combinations 
of the electron and antiproton. The fit
is required to have a $\chi^2$ per degree of freedom less than 10.


\section{Partial Reconstruction Technique}

\begin{figure}
\includegraphics[width=4.0in]
{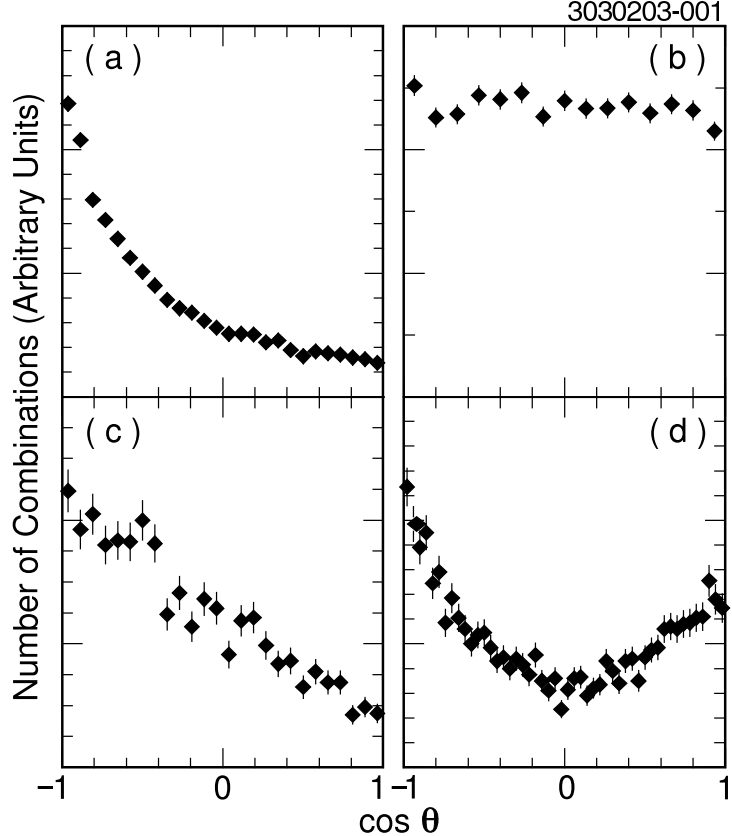}
\caption{Distribution of the cosine of the angle between 
same sign electrons and antiprotons ($\cos\theta$). 
Plot (a) shows $\ep$ signal combinations from $\btoalcpen$ decay; 
plot (b) shows uncorrelated background; 
plot (c) shows correlated background.  
Plots (a), (b), and (c) are obtained using the 
CLEO $\BBbar$ Monte Carlo generator. 
Plot (d) shows continuum backgrounds obtained from data.
}
\label{fig:shapes}
\end{figure}

We study the angular correlations between the prompt 
electron and antiproton.  If we define $\theta$ as the angle between the 
electron and the antiproton, the corresponding 
cos($\theta$) distribution is peaked at $\cos(\theta)=-1$ (back-to-back)
for signal events.  
Figure~\ref{fig:shapes} 
shows the $\cos(\theta)$ distributions for $\btoalcpen$ signal events and 
various backgrounds.  We will use the difference between the signal and 
background shapes in this distribution to fit for the amount of signal 
in our sample.

There are four main sources of backgrounds as follows:

\begin{itemize}
\item  Uncorrelated background:\\
This includes the e/{$\overline p$} combinations where the electron and 
antiproton are from opposite $B$ meson decays 
(see Figure~\ref{fig:shapes}(b)). 
The $\cos(\theta)$ distribution of this background is almost flat, modulo 
a fiducial acceptance correction as seen from Monte Carlo.
 
\item Correlated background:\\
This includes non-prompt e/{$\overline p$} combinations, 
which are from the same 
$B$ meson but not from a 
signal event, such as in the decay chain: 
$B^{+}\rightarrow \Lambda_{c}^-X$, 
$\Lambda_{c}^- \rightarrow \overline \Lambda e^-X$, 
$\overline \Lambda\rightarrow \overline{p}X$ (see Figure~\ref{fig:shapes}(c)). 
The $\ctt$ distribution of this background as found from Monte Carlo 
is also peaked near 
$\cos(\theta)\simeq -1$, but less sharply than signal.

\item Continuum background:\\
This is the background due to non-$\BBbar$ sources, 
\textit{i.e.} $e^+e^-\rightarrow q \overline q$, 
where $q = u,d,s,c$ (see Figure~\ref{fig:shapes}(d)) found using 
data collected at energies below the $\Upsilon(4S)$.  
 
\item Fake e/{$\overline p$} background:\\
This is due to particles misidentified as electrons or antiprotons and 
is found using data.

\end{itemize}

We obtain the overall $\ep$ angular distributions, 
\textit{i.e.} $\cos(\theta)$ 
distributions between electrons and antiprotons, for each of the 
CLEO~II and CLEO~II.V datasets separately and then combine them.  
The $\ep$ angular distribution found from the off-resonance data sample 
is scaled by 
luminosity and the energy dependent four-flavor cross section and 
then subtracted (the scale factor is approximately 2).
We subtract the fake electron and antiproton 
backgrounds using data distributions as described below.  After 
these subtractions, the angular distribution is composed of uncorrelated 
background, correlated background, and possibly signal.  Using Monte 
Carlo generated shapes for each of these contributions, we fit to 
a sum of these three components to determine the yield of the signal events. 
Table~\ref{tab:data} gives the overall yields for the two data samples.

\begin{table*}
\caption{Yields of events from the CLEO~II and CLEO~II.V data samples,  
integrated over the entire angular 
distribution. The last row shows the yield 
after subtracting the continuum and fake backgrounds.}
\label{tab:data}
\medskip
\begin{tabular}{|c|c|c|}
\hline
Event Type  & CLEO~II  & CLEO~II.V \\ \hline
$B\overline{B}$ Events  & $3,328,000\pm 67,000$& $6,372,000\pm 127,000$\\
Overall $\ep$ Combinations & $10193\pm 101$ & $16829\pm 130$\\ 
Continuum background (scaled)   & $3656\pm 84$ & $6471\pm 114$ \\
Fake $e$ background      & $212\pm 40$ & $308\pm 58$ \\
Fake $p$ background      & $1872\pm 159$ &  $2859\pm 243$\\
\hline
Background subtracted distribution  & $4453\pm 210$ & $7191\pm 304$\\
\hline
\end{tabular}
\end{table*}


The subtractions of the misidentified electron and misidentified 
antiproton backgrounds follow similar procedures, described 
here for the fake electrons.  The 
fake electron angular distribution is found using the following equation: 
\begin{displaymath}
fbkgd(\theta) = \sum_{p=0.6}^{1.5}\sum_{i}fdist(\cos(\theta), p) 
\times misid_{i,p}.\qquad 
\end{displaymath}
Here $\cos(\theta)$ is the angle between the antiproton 
and fake electron, $p$ is 
the momentum of the fake electron (in GeV/$c$), 
$i =\pi,K,p,\mu$ ; \textit{fbkgd} is the 
$\ctt$ distribution of $\ep$ combinations 
that contain a fake electron, \textit{i.e.} the fake electron background;   
\textit{$fdist$} is the angular distribution of non-electrons in each 
momentum range (obtained by processing data with an electron 
anti-identification cut); and \textit{$misid_{i,p}$} is the electron 
misidentification probability as a function of momentum,
which is calculated by multiplying the abundance of 
each particle species (found in Monte Carlo) by its corresponding electron 
misidentification rate (obtained from data) in each momentum range.
The electron and positron misidentification probabilities 
are less than 0.3\% per track so there is very 
little background from this source.  
The proton and antiproton misidentification probabilities range from 
0.2\% per track at lower momenta to 3\% per track at higher momenta.   
The statistical error associated with particle abundance and misidentification
rates is determined by the data and Monte Carlo sample sizes, and included in the 
statistical error from the fit to the final $\ep$ angular distribution. 


We use the CLEO $\BBbar$ Monte Carlo to obtain the uncorrelated and 
correlated background angular distribution shapes.
For the signal, the angular distribution shapes as well as the efficiency of 
our event selection are found using the standard CLEO Monte Carlo event 
generator as well as a phase space generator. The CLEO Monte Carlo generator 
(hereafter referred to as ``V-A model'') generates a decay such as 
$\btoalcpen$ in two steps. 
The first step is the semileptonic 
decay of $b\rightarrow cW, W\rightarrow \ell \overline \nu_{\ell}$, preserving 
the V-A structure of the weak decay.  
This step involves a three body decay, with three 
initial particles produced: $e^-, \overline \nu_{e}$ and a 
($\Lambda_c \overline p$)
pseudo-particle.  At the second step, 
the pseudo-particle decays 
into two particles: $\Lambda_c$ and $\overline p$, ignoring 
any possible spin correlation.
The same mechanism is used to generate the other decay modes, 
the only difference being that the intermediate state 
pseudo-particle in the V-A model is varied.  The phase space 
model used is simply a four-body $B$ decay, with all the final state particles 
generated at one step.  The subsequent CLEO detector 
simulation is GEANT based~\cite{geant}.

\begin{figure*}
\includegraphics[width=6.0in]
{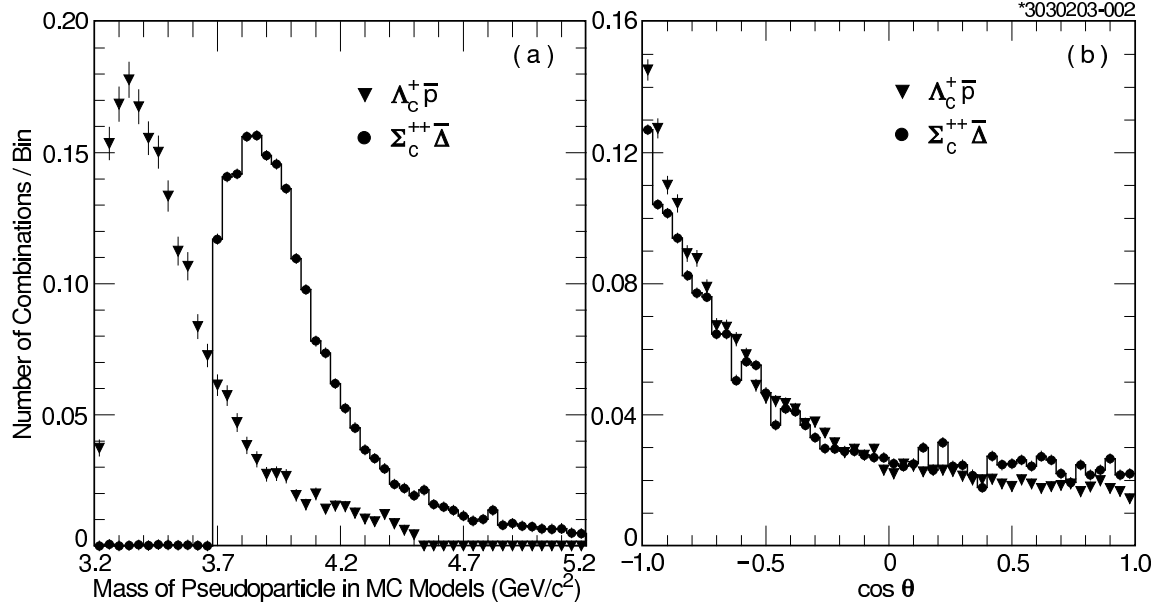}
\caption{Comparison of signal Monte Carlo 
models for $\btolcpen$ and 
$\overline B^0\rightarrow\Sigma_{c}^{++}\overline
\Delta^{-} e^-\overline \nu_{e}$. 
Plot (a) displays the invariant mass of 
pseudo-particle~($\Lambda_c \overline{p}$/
$\Sigma_{c}^{++}\overline\Delta^{-}$). 
Plot (b) displays the $\cos(\theta)$ distributions of $\ep$ combinations.
The black trangles show the expectations for the $\btolcpen$ decay and  
the histogram with error bars shows the 
$\overline B^0\rightarrow\Sigma_{c}^{++}\overline\Delta^{-} 
e^-\overline \nu_{e}$ decay mode.  For the sake of 
comparison, the distributions have been normalized to unit area.}
\label{figure:COMP_SIGNAL_B2C_1PION}
\end{figure*}

In the V-A model, the mass of the 
pseudo-particle could affect the 
angular distribution between $e$ and $\overline p$ and the electron 
and antiproton momentum distributions. In the 
standard CLEO Monte Carlo 
event generator, the mass spectrum of the 
pseudo-particle~($\Lambda_c \overline p$)
is generated as a phase space modified Breit Wigner distribution, 
with a central mass of 
3.35 GeV/$c^2$, and a width of 0.50 GeV/$c^2$,  
as shown in 
Figure~\ref{figure:COMP_SIGNAL_B2C_1PION}(a).  
This  pseudo-particle~($\Lambda_c \overline p$)
mass spectrum reproduces the measured inclusive 
$B \rightarrow \Lambda_c X$ and 
$B \rightarrow p X$ momentum spectra~\cite{pspec}.  In order to allow the 
possibility of a lower efficiency, we examine two-body decays into the 
baryon/antibaryon system $X_c\overline{N}$.  
We have analyzed the 
$\cos(\theta)$ distributions from the following decay modes:
$B^-\rightarrow\Lambda_{c}^+\overline{p} e^-\overline \nu_{e}$,
$B^-\rightarrow\Sigma_{c}^{+}\overline p e^-\overline \nu_{e}$,
$\overline B^0\rightarrow\Sigma_{c}^{++}\overline p e^-\overline \nu_{e}$,
$B^-\rightarrow\Sigma_{c}^{++}\overline\Delta^{--} e^-\overline \nu_{e}$,
$\overline B^0\rightarrow\Sigma_{c}^{++}\overline\Delta^{-} e^-\overline \nu_{e}$,
$B^-\rightarrow\Sigma_{c}^{0}\overline\Delta^{0} e^-\overline \nu_{e}$,
$B^-\rightarrow\Sigma_{c}^{+}\overline\Delta^{-} e^-\overline \nu_{e}$ and
$\overline B^0\rightarrow\Sigma_{c}^{+}\overline\Delta^{0} e^-
\overline \nu_{e}$.  The decay mode
$\overline B^0\rightarrow\Sigma_{c}^{++}\overline
\Delta^{-} e^-\overline \nu_{e}$ 
provides the softest lepton momentum spectrum and 
therefore the smallest efficiency 
for this analysis ($13.5\pm 0.2$)\%.  The efficiency is calculated 
for modes with a $\overline{p}$ in the final state.    
The efficiency from the decay mode $\btolcpen$ 
is is the highest at ($20.7\pm 0.1$)\%.  For comparison, the 
pseudo-particle ($\Sigma_{c}^{++}\overline\Delta^{-}$) mass spectrum which 
was generated with a central mass of 3.85 GeV/$c^2$, 
a width of 0.50 GeV/$c^2$, and a 
threshold mass of 3.68 GeV/$c^2$, is 
also shown in Figure~\ref{figure:COMP_SIGNAL_B2C_1PION}(a).  
Figure~\ref{figure:COMP_SIGNAL_B2C_1PION}(b) shows the 
angular distribution of signal $\ep$ 
combinations for the two modes.   For the signal model, 
we combine these two modes in equal ratios and bracket the model dependence 
by choosing a model with 100\% of either of the two decay modes. 

\begin{figure*}
\includegraphics[width=6.0in]
{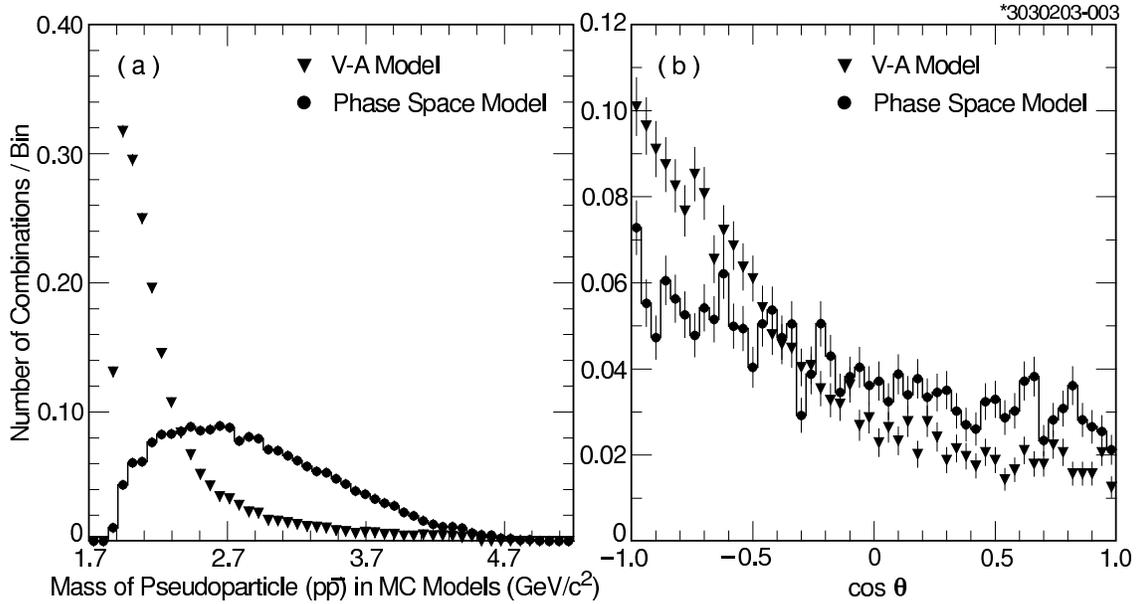}
\caption{Comparison of two signal Monte Carlo 
models for $\btoppen$ decay. 
Plot (a) displays the invariant mass of pseudo-particle~($p \overline p$). 
Plot (b) displays the  
$\cos(\theta)$ distributions of $\ep$ combinations for 
the 2 models considered. 
The black triangles show the expectations from  
the V-A model, while the histogram shows the 
expected distribution for the phase space model.  
For the sake of comparison, the distributions have 
been normalized to unit area.}
\label{figure:COMP_SIGNAL_B2U}
\end{figure*}

Figure~\ref{figure:COMP_SIGNAL_B2U} 
compares the V-A and phase space models for the $\btoppen$ decay mode. 
It shows that the two Monte 
Carlo models give significantly different angular distributions for the $\ep$ 
combinations in this decay.  We choose the phase space model to bracket the 
possible efficiencies and angular distributions of various models.

\section{Search for $\btoc$ decays}

\begin{figure}
\includegraphics[width=4.0in]
{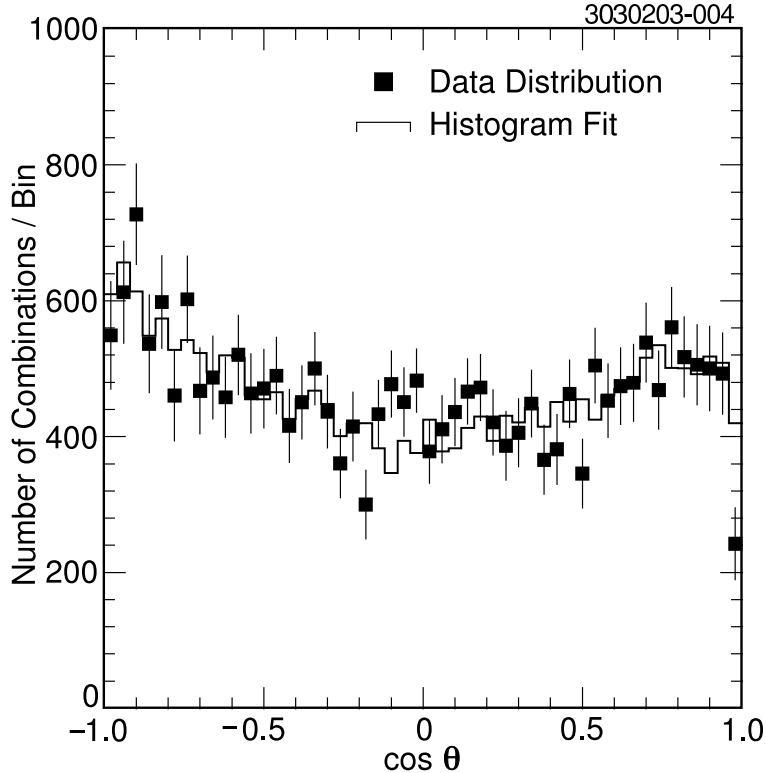}
\caption{The $\cos(\theta)$ distributions found in data after 
subtracting the continuum, fake electron, and fake 
antiproton backgrounds.  
The plot shows the fit to the combined CLEO~II and CLEO~II.V datasets using 
Monte Carlo distributions for the $b\rightarrow c$ signal 
(as discussed in the text), 
correlated background and uncorrelated background.
The confidence level of the fit is 29\%.}
\label{figure:fit_b2c}
\end{figure}

The $\cos(\theta)$ distributions for $\ep$ combinations after 
subtracting the continuum, fake electron, and antiproton backgrounds 
are shown in Figure~\ref{figure:fit_b2c} along with the results of the fit. 
In the fit, we use the shapes obtained from Monte Carlo 
(Figure~\ref{fig:shapes}(a)(b)(c) and 
Figure~\ref{figure:COMP_SIGNAL_B2C_1PION}(b)) 
and allow each of the normalizations of the three components to float 
independently.  
Table~\ref{tab:btoc} gives the results from the fit.   
There is no evidence for a signal so we calculate an upper limit.  
From the fit we find 
${\cal B} (\btoxpen)=(2.5 \pm 1.9\pm 1.1\pm 1.4) \times 10^{-4}$, 
corresponding to a 90\% $C.L.$ upper limit of 
${\cal B} (\btoxpen) < 5.9 \times 10^{-4}$. The last error is 
the model dependence error found from varying the composition of 
light-mass states with higher resonance states. 

\begin{table*}
\caption{ Results from the fits for the $\btoxpen$ analysis using a 50\% - 50\%
mix of two decay modes: $\btolcpen$ and 
$\overline B^0\rightarrow\Sigma_{c}^{++}\overline\Delta^{-} e^-
\overline \nu_e$. 
The first row shows the number of signal events found, with the statistical 
error determined from the fit and systematic errors 
determined as discussed in the text.  
The second and third rows show correlated and uncorrelated 
backgrounds from the fit, 
respectively. The result is presented with the statistical, systematic, and 
model dependence errors in the sixth row.  These errors are 
combined in quadrature to obtain the upper limit listed in the last row.}
\label{tab:btoc}
\medskip
\begin{tabular}{|c|c|}
\hline
Event Type & Events \\ \hline 
Signal events (fit) & $834\pm634\pm380$ \\ 
Correlated background (fit)& $-331\pm1729$ \\ 
Uncorrelated background (fit) & $11141\pm1303$ \\ 
Avg. Efficiency from Monte Carlo~& $(17.1\pm0.1)\%$ \\ 
Efficiency corrected data & $4877\pm 3708\pm2224$ \\ \hline
${\cal B} (\btoxpen)$ & $(2.5\pm1.9\pm1.1\pm 1.4)\times 10^{-4}$ \\
Upper Limit of ${\cal B}$ (90\% $C.L.$) & $5.9\times 10^{-4}$ \\
\hline
\end{tabular}
\end{table*}

\begin{table}
\caption{Systematic errors for the measurement of $\btoxpen$.  These are the 
contributions to the systematic error listed on the first line of 
Table~\ref{tab:btoc}}
\label{tab:btoc_cleo2_syserror}
\medskip
\begin{tabular}{|c|c|}
\hline
Systematic Error & Events \\ \hline
Correlated background & $\pm  98$ \\ 
Uncorrelated background & $\pm 183$ \\
Fake proton background subtraction & $\pm 299$ \\
Fake electron background subtraction & $\pm  29$ \\
Proton identification efficiency & $\pm  75$ \\ 
Electron identification efficiency & $\pm  25$ \\ 
Vertex constrained fit efficiency & $\pm  63$ \\
Signal Monte Carlo sample statistics & $\pm  33$ \\ \hline
Total & $\pm 380$ \\
\hline
\end{tabular}
\end{table}

Table~\ref{tab:btoc_cleo2_syserror} summarizes the systematic errors. 
The systematic errors include those associated with each of the 
backgrounds: correlated, uncorrelated, fake proton and fake electron, as 
described in more detail below.  The two largest errors come from the fake 
proton subtraction and variations allowed in the uncorrelated background.

The correlated 
background (Figure~\ref{fig:shapes}(c)) has a similar shape to that 
of the signal.  To calculate a conservative systematic error from this source, 
we refit the data assuming no correlated background exists and take 
the difference between the central value in this fit and the original. 

The uncorrelated background systematic error is found from a combination 
of normalization and shape errors.  If we assume there is no signal or 
correlated background, we can 
scale the Monte Carlo normalization by the number of events 
and compare it with the 
data.  There are a total of 16\% fewer data events than in the scaled Monte 
Carlo; we use this difference to account for the normalization error.  
The angular 
distribution of the uncorrelated background is expected to 
be flat in the absence of acceptance effects (see Figure~\ref{fig:shapes}(b)). 
However, as we only accept tracks in the barrel region of the detector, \textit{i.e.} 
$\vert\cos(\theta _{dip})\vert<0.71$, the $\ep$ combinations passing the cuts have 
slightly higher probability to come from the two opposite barrel regions.  
Therefore, the Monte Carlo angular distribution of this background 
is peaked towards $\cos(\theta)\simeq \pm 1$.  Because of finite spatial 
seqmentation effects, two tracks very close together have a slightly lower 
efficiency than those 
that are more back-to-back diminishing the peak near $\cos(\theta)=1$. 
We change the shape in the uncorrelated background to a symmetric distribution 
and fit again; the difference in the fitted central values is 30\%.  
We take half 
of this ``shape'' difference (15\%) and combine it in quadrature with the 
normalization difference to find an overall systematic error for the 
uncorrelated background of 22\%. 

\begin{figure*}
\includegraphics*[width=6.0in]
{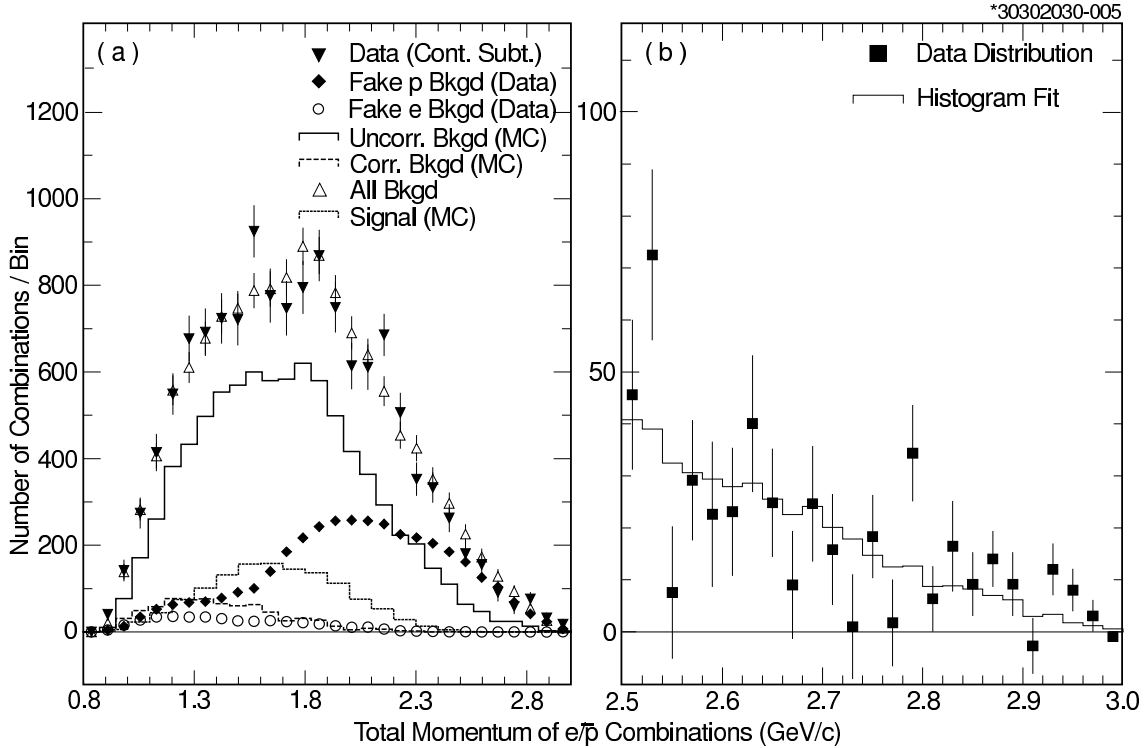}
\caption{CLEO~II Data/Monte Carlo $\pepp$ distribution. 
Plot (a) shows 
the total momentum sum of the electron and 
antiproton tracks, from different data and Monte Carlo components. 
The components include: 1) fake proton background from data (black diamonds); 
2) fake electron background from data (asterisks); 
3) uncorrelated background from Monte Carlo (solid line), and  
4) correlated background from Monte Carlo (dashed line). 
The outermost empty triangles represent the sum of all the above backgrounds. 
The filled black triangles show the overall data distribution, 
with the continuum background subtracted. 
Plot (b) is the fit to the final data distribution 
(continuum and uncorrelated background subtracted), 
using the fake proton background
distribution in the region above 2.5~$\GeVc$.}
\label{figure:fakepbkgd_error_cleo2_ss}
\end{figure*}

We study additional systematic errors from the fake proton background subtraction by 
comparing the $\pepp$ distribution in data and Monte Carlo.
Figure~\ref{figure:fakepbkgd_error_cleo2_ss} shows that in the 
$\pepp$ region above 2.5~$\GeVc$, the backgrounds remaining are limited to 
the fake proton and 
the uncorrelated background. A Monte Carlo study shows that there are no 
$\btolcpen$ signal events in this region in any scenario. 
The fake electron background is very small compared to the fake proton 
background as seen in Figure~\ref{figure:fakepbkgd_error_cleo2_ss}(a).  
Therefore, in the region 
above $(2.5-3.0)$~$\GeVc$, if we
use the scaled Monte Carlo to subtract the uncorrelated background, 
the remaining $\pepp$ data distribution should be saturated by the 
predicted fake proton background (as shown in 
Figure~\ref{figure:fakepbkgd_error_cleo2_ss}(b)).  We 
estimate the systematic error from the fake proton background subtraction from 
the deviation from complete saturation.
 The fit gives a difference in normalization of 
$\sim $15\% between the amount of predicted fake proton background and that 
obtained for the best fit to the data, which implies that the fake proton 
background may be systematically wrong by $\sim $15\%. We then shift 
the fake antiproton background normalization 
by $\pm15\%$ and redo the fit to the final $\ep$ angular distribution. 
The difference between the central values obtained from the new fit vs. the 
original fit is taken as the systematic error for the fake antiproton 
background subtraction.  For the systematic error from misidentified electrons, 
studies using real pions and kaons in data have been done which determine the 
errors on the fake probabilities.  These fake probability errors and the error 
associated with using an antielectron identification cut for counting tracks in the 
data are folded together to combine for an estimate of $\pm20\%$ from this source.  
This technique is confirmed using a Monte Carlo test 
which verifies that the number of misidentified particles calculated is 
consistent with the number generated, and that a 20\% error is a 
conservative estimate.  
To calculate the effect on our data sample, we shift the fake electron background 
normalization by $\pm20\%$, redo the fits and take the difference between the 
new fit and the original fit as the systematic error from this source.

In addition, errors are added to account for uncertainties in the 
antiproton and electron 
identification efficiency differences between Monte Carlo and data.   
The antiproton identification efficiency is found using an antiproton 
data sample from $\overline{\Lambda}\rightarrow\overline{p}\pi$ in continuum 
data, as a 
function of momentum.  The momentum spectrum for protons in our Monte Carlo 
signal sample is used to weight these efficiencies.  The overall 
error from this source is estimated to be 9\%.  Similarly, 
for electrons, a CLEO study using radiative Bhabha events in the data itself has 
determined an overall error of 3\%.  

The error from the continuum background subtraction is statistical,  
determined by the size of the data sample, and is directly incorporated 
into the final statistical error, as is the statistical error due to the 
limited Monte Carlo sample size.  There is also an error due to the 
systematics associated with the constrained vertex fit.  This is taken to 
be half of the inefficiency found from the signal Monte Carlo sample with and 
without the cut (7.5\%).


\section{Search for the $\btou$ decay $\btoppen$} 

\begin{figure}
\includegraphics[width=4.0in]
{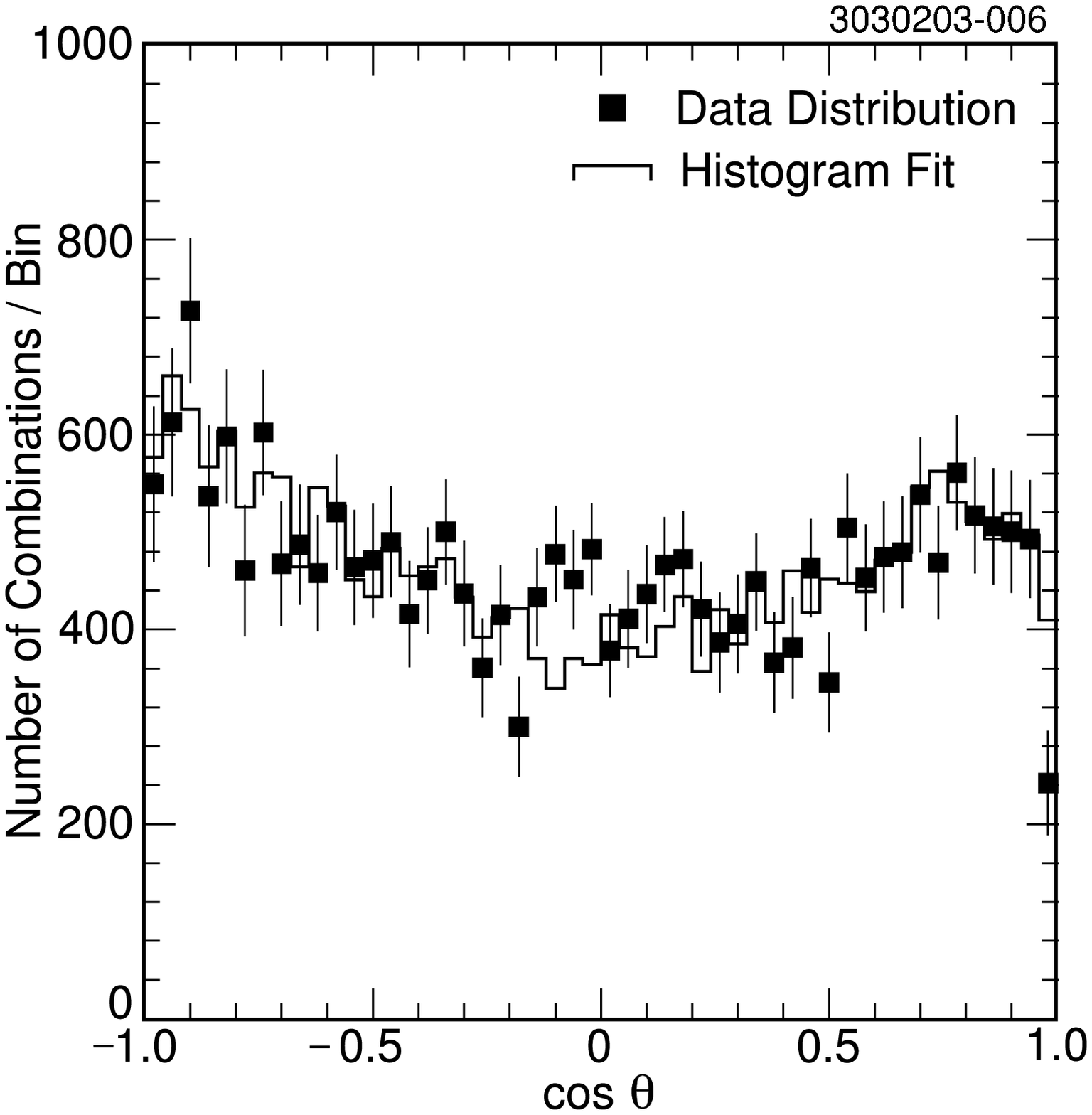}
\caption{The $\cos(\theta)$ distributions found in data after 
subtracting the continuum, fake electron, and fake antiproton backgrounds. 
The plot shows the fit using a $b\rightarrow u$ signal model ($\btoppen$).   
The confidence level for the fit is 34.5\%.
Note that the only difference between this Figure and 
Figure~\ref{figure:fit_b2c} 
is the simulated signal shape.}
\label{figure:fit_b2u}
\end{figure}

\begin{table*}
\caption{ Results from the fits for the $\btoppen$ analysis using the V-A model.  
The first row shows the number of signal events found, with the statistical 
error determined from the fit and systematic errors 
determined as discussed in the text.  
The second and third rows show correlated and uncorrelated backgrounds from the fit, 
respectively. ``Efficiency corrected data'' are results found using the V-A signal 
Monte Carlo generator model. 
The statistical and systematic errors are combined in quadrature 
for the final result.}
\label{tab:btou_1}
\medskip
\medskip
\begin{tabular}{|c|c|}
\hline
Event Type & CLEO~II and CLEO~II.V datasets \\ \hline 
Signal events (fit) & $1685\pm1068\pm1032$ \\ 
Correlated background (fit) & $-2665\pm2937$ \\ 
Uncorrelated background (fit) & $12624\pm1991$ \\ 
Efficiency from Monte Carlo & $(14.9\pm0.2)\%$ \\ 
Efficiency corrected data & $11309\pm 7169\pm 6930$ \\ \hline
${\cal B} (\btoppen)$ & $(5.8\pm3.7\pm3.6)\times 10^{-4}$ \\
Upper Limit of ${\cal B}$ (90\% $C.L.$) & $1.2\times 10^{-3}$ \\
\hline
\end{tabular}
\end{table*}

We can also fit the $\ep$ angular distribution to the $b\rightarrow u$ 
signal decay channel $\btoppen$. 
Figure~\ref{figure:COMP_SIGNAL_B2U} shows that the two Monte Carlo generator 
models give quite different signal $\ep$ angular distributions for
this decay mode.
Figure~\ref{figure:fit_b2u} shows the fits to the 
CLEO~II and CLEO~II.V $\ctt$ distributions,  
assuming signal events are entirely from $\btoppen$ decay, where the signal 
Monte Carlo events are obtained using the V-A model generator.  
We see no evidence for a $b\rightarrow u$ signal from this decay mode.  
Table~\ref{tab:btou_1} gives the results based on the V-A model.  
Systematic errors are calculated using the same procedures described 
above, for the $b\rightarrow c$ analysis.  We obtain the branching ratio 
${\cal B} (\btoppen)=(5.8 \pm 3.7\pm 3.6) \times 10^{-4}$, 
corresponding to a 90\%~$C.L.$ upper 
limit of ${\cal B} (\btoppen) < 1.2 \times 10^{-3}$.  
For the phase space model, 
combining the CLEO~II and CLEO~II.V datasets, we obtain a branching ratio of  
${\cal B} (\btoppen)=(2.6 \pm 1.1\pm 1.6) \times 10^{-3}$, corresponding to 
an upper limit of ${\cal B} (\btoppen) < 5.2 \times 10^{-3}$ (90\% $C.L.$).


\section{\bf Conclusion}

The angular distribution between electrons and antiprotons has 
been studied to search for semileptonic baryon decays from $B$ mesons.  
The analysis was optimised to search for the $b\rightarrow c$ decay
$\btolcpen$.  For the $b\rightarrow c$ modes, we use a (50\%-50\%) mixture 
of $\btolcpen$ and $\overline{B}^0\rightarrow\Sigma_C^{++}
\overline\Delta^{-}e^-\nu_e$ signal modes and perform a fit to the 
angular distribution. We see no evidence for a signal and measure an 
upper limit at  90\% $C.L.$, combining the CLEO~II and CLEO~II.V 
data samples together, of 
$${\cal B} (B\rightarrow\overline{p}e^-\overline \nu_{e}X) < 
5.9\times 10^{-4} \quad ({\rm V-A~model}).$$ 

These results are an improvement upon the previous limits~\cite{lin,arg2}, 
in support of
their conclusion that the semileptonic decay of $B$ mesons into 
baryons is not large 
enough to cover the discrepancy in the $B$ meson semileptonic 
branching ratio between
theoretical prediction and experimental measurements~\cite{argus,theory1}.  
In particular, these results show that charmed baryon production in 
semileptonic $B$ decay is less than 1.2\% of all semileptonic $B$ decays, 
as compared with $\Lambda_C$ production in generic $B$ decays at 
$(6.4\pm 1.1)$\%~\cite{PDG}.  The results also suggest 
that the dominant mechanism 
for baryon production in generic $B$ decays is not external $W$ emission.

We also searched for the $b \rightarrow u$ decay $\btoppen$.  
We obtain the following upper limits at 90\% $C.L.$ for each of the models:
\begin{tabular}{r @{$~<~$} l}
${\cal B}(\btoppen)$ & $1.2\times 10^{-3} \quad ({\rm V-A})$\\
                   & $5.2\times 10^{-3} \quad ({\rm phase~space}).$
\end{tabular}
These limits do not constrain any theories at this time.

\section{\bf Acknowledgements}

We gratefully acknowledge the effort of the CESR staff in providing us 
with excellent
luminosity and running conditions. M.~Selen thanks the Research Corporation,  
and A.~H.~Mahmood thanks the Texas Advanced Research Program. 
This work was supported by the National 
Science Foundation and the U.S. Department of Energy.


\end{document}